\documentclass[final,5p,times,twocolumn,nopreprintline]{elsarticle}
\usepackage{amsmath,amssymb,amsfonts,dsfont}
\usepackage{graphicx,color}
\usepackage{slashed}
\usepackage{booktabs,tabulary}

\newcommand{\Fpi}{F_\pi}
\newcommand{\mpi}{M_\pi}
\newcommand{\ga}{g_A}
\newcommand{\Order}{\mathcal{O}}
\newcommand{\Op}{\mathcal{O}}
\newcommand{\mN}{m_N}
\newcommand{\eps}{\varepsilon}
\newcommand{\mD}{m_\Delta}
\newcommand{\MeV}{\,\text{MeV}}
\newcommand{\GeV}{\,\text{GeV}}
\newcommand{\Lamb}{\Lambda_\text{b}}
\newcommand{\Lagr}{\mathcal{L}}
\newcommand{\Nc}{N_c}
\newcommand{\beq}{\begin{equation}}
\newcommand{\eeq}{\end{equation}}
\newcommand{\bra}[1]{\langle{#1}\vert }
\newcommand{\ket}[1]{\vert {#1}\rangle }
\newcommand{\clebsch}[6]{\left(
\left.
\begin{array}{ccc}
{#1}&{#2}\\
{#4}&{#5}
\end{array}\right|
\begin{array}{c}
{#3}\\
{#6}
\end{array}\right)}

\newcommand{\toright}[1]{\hspace*{\fill}{\footnotesize{#1}}}

\allowdisplaybreaks[1]

\begin{document}

\renewcommand{\theequation}{\arabic{equation}}

\begin{frontmatter}

\title{\toright{\textnormal{INT-PUB-16-035, NSF-KITP-16-158
}}\\Reconciling threshold and subthreshold expansions for pion--nucleon scattering}

\author[Bochum]{D.\ Siemens}
\author[HISKP,Bern]{J.\ Ruiz de Elvira}
\author[Bochum,KITP]{E.\ Epelbaum}
\author[INT,KITP]{M.\ Hoferichter}
\author[Bochum,KITP]{H.\ Krebs}
\author[HISKP]{B.\ Kubis}
\author[HISKP,Julich]{U.-G.\ Mei{\ss}ner}

\address[Bochum]{Institut f\"ur Theoretische Physik II, Ruhr-Universit\"at Bochum, D--44780 Bochum, Germany}
\address[HISKP]{Helmholtz--Institut f\"ur Strahlen- und Kernphysik (Theorie) and
   Bethe Center for Theoretical Physics, Universit\"at Bonn, D--53115 Bonn, Germany}
\address[Bern]{Albert Einstein Center for Fundamental Physics, Institute for Theoretical Physics,
University of Bern, Sidlerstrasse 5, CH--3012 Bern, Switzerland}
\address[KITP]{Kavli Institute for Theoretical Physics, University of California, Santa Barbara, CA 93106, USA}
\address[INT]{Institute for Nuclear Theory, University of Washington, Seattle, WA 98195-1550, USA}
\address[Julich]{Institut f\"ur Kernphysik, Institute for Advanced Simulation, and 
   J\"ulich Center for Hadron Physics, Forschungszentrum J\"ulich, D--52425  J\"ulich, Germany}

\begin{abstract}
 Heavy-baryon chiral perturbation theory (ChPT) at one loop fails in relating the pion--nucleon amplitude in the physical region and for subthreshold kinematics due to loop effects enhanced by large low-energy constants.
 Studying the chiral convergence of threshold and subthreshold parameters up to fourth order in the small-scale expansion, we address the question to what extent this tension can be mitigated by including the $\Delta(1232)$ as an explicit degree of freedom and/or using a covariant formulation of baryon ChPT. 
 We find that the inclusion of the $\Delta$ indeed reduces the low-energy constants to more natural values and thereby improves consistency between threshold and subthreshold kinematics.
 In addition, even in the $\Delta$-less theory the resummation of $1/\mN$ corrections in the covariant scheme improves the results markedly over the heavy-baryon formulation, 
 in line with previous observations in the single-baryon sector of ChPT that so far have evaded a profound theoretical explanation.
\end{abstract}

\begin{keyword}
Pion--baryon interactions\sep Dispersion relations\sep Chiral Lagrangians\sep Chiral symmetries

\PACS 13.75.Gx\sep 11.55.Fv\sep 12.39.Fe\sep 11.30.Rd
\end{keyword}

\end{frontmatter}

\section{Introduction}
\label{sec:intro}

The approximate chiral symmetry of QCD imposes strong constraints on low-energy hadron dynamics,
which can be explored systematically in the framework of chiral perturbation theory (ChPT)~\cite{Weinberg:1978kz,Gasser:1983yg,Gasser:1984gg}.
While in the meson sector the expansion proceeds directly in terms of momenta and quark masses divided by a breakdown scale $\Lamb$, typically identified 
with the mass of the $\rho(770)$ or the scale of chiral symmetry breaking $\Lambda_\chi=4\pi\Fpi\sim1.2\GeV$, in the baryon sector the nucleon mass $\mN$ represents a new scale that needs to
be taken into account in order not to spoil the chiral power counting~\cite{Gasser:1987rb}. 
Heavy-baryon ChPT (HBChPT)~\cite{Jenkins:1990jv,Bernard:1992qa} achieves this by systematically expanding the effective Lagrangian in $1/\mN$, identifying $\Lamb\sim\mN$.
In subsequent years, several variants of covariant baryon ChPT have been developed~\cite{Ellis:1997kc,Becher:1999he,Goity:2001ny,Schindler:2003xv,Schindler:2003je,Bernard:2007zu},
in which the power-counting-violating part is subtracted in one way or another.
While originally motivated by the desire to preserve the analytic structure of the amplitude in the vicinity of anomalous thresholds and unitarity cuts, it has also been observed 
that the resummation of $1/\mN$ corrections can improve the phenomenology even in kinematic regions where the HB formulation does reproduce the analytic structure correctly~\cite{Kubis:2000zd,Becher:2001hv,Alarcon:2012kn,Siemens:2016hdi}.

The efficacy of different formulations of baryon ChPT has implications beyond the single-nucleon sector. In chiral effective field theory, the extension of ChPT to multi-nucleon systems~\cite{Weinberg:1990rz,Weinberg:1991um,Ordonez:1995rz,Epelbaum:2008ga,Machleidt:2011zz}, the low-energy constants (LECs) that appear in pion--nucleon ($\pi N$) scattering determine the long-range part of the nucleon--nucleon ($NN$) potential as well as three-nucleon forces. While the use of the HB formulation is common to all implementations to date, $1/\mN$ corrections are often 
counted suppressed by one additional order compared to the standard single-nucleon HB counting, to account for the fact that the breakdown scale in the multi-nucleon sector tends to be lower than in single-nucleon applications~\cite{Weinberg:1991um,Epelbaum:2008ga} (this counting scheme will be referred to as HB-$NN$ counting in the following, in contrast to the standard HB-$\pi N$).

Recently, the combination of dispersion theory in the form of Roy--Steiner (RS) equations~\cite{Ditsche:2012fv,Hoferichter:2012wf,Hoferichter:2015dsa,Hoferichter:2015tha,Hoferichter:2015hva,Hoferichter:2016ocj,Hoferichter:2016duk} with precision measurements of the $\pi N$ scattering lengths in pionic atoms~\cite{Strauch:2010vu,Hennebach:2014lsa,Gotta:2008zza,Baru:2010xn,Baru:2011bw} 
resulted in a reliable representation of the $\pi N$ scattering amplitude in the whole low-energy region, both in the physical region and for subthreshold kinematics.
Surprisingly, the matching to HBChPT revealed that, in contrast, the chiral representation is not accurate enough to relate these two regions~\cite{Hoferichter:2015tha}.
These findings can be best illustrated considering the parameters in the expansion around threshold and the subthreshold point: with LECs determined in the subthreshold region, where due to 
the absence of unitarity cuts ChPT is expected to converge best~\cite{Colangelo:2001df}, the chiral series fails to reproduce some of the threshold parameters.
The reason for this behavior can be traced back to loop diagrams producing terms that scale as $\ga^2(c_3-c_4)\sim -16\GeV^{-1}$, an enhancement that is, at least partially, generated by
 saturation of the LECs $c_i$ with the $\Delta(1232)$ resonance. As argued in~\cite{Hoferichter:2015tha}, this inconsistency between subthreshold and threshold kinematics implies that
 in a HB formulation, LECs determined at the subthreshold point are preferable for multi-nucleon applications, given that the kinematics for the two-pion exchange in the $NN$ potential are much closer to the subthreshold point than to the physical region in $\pi N$ scattering.
 
In this paper we address the question to what extent consistency between subthreshold and physical region can be restored by introducing the $\Delta$ as an explicit degree of freedom, and/or by using a covariant formulation of baryon ChPT. The $\Delta$ is included within the small-scale expansion~\cite{Hemmert:1997ye}, counting the difference $\eps=\mD-\mN$ in the same way as a momentum scale $p$.  
$\pi N$ scattering with explicit $\Delta$ degrees of freedom has been considered before at $\Order(\eps^3)$ in HB~\cite{Fettes:2000bb} and covariant~\cite{Yao:2016vbz} formulations, as well as within the $\delta$-counting of~\cite{Pascalutsa:2002pi} up to $\Order(p^3)$ in a covariant scheme~\cite{Alarcon:2012kn} (see also~\cite{Chen:2012nx}). 
Here, we extend the analysis to full one-loop order $\Order(\eps^4)$ and study the predictions for the leading eight threshold parameters, with LECs determined from the subthreshold parameters predicted by the RS analysis~\cite{Hoferichter:2015hva}. After a brief introduction to the formalism in Sect.~\ref{sec:formalism}, we first present the results when including the $\Delta$ in HBChPT in Sect.~\ref{sec:Delta}, and then extend the analysis towards a covariant formulation in Sect.~\ref{sec:covariant}. We offer our conclusions in Sect.~\ref{sec:conclusions}.
Details on large-$\Nc$ constraints and correlation coefficients of the extracted LECs are summarized in the appendices.

\section{Formalism}
\label{sec:formalism}

For the calculation of the threshold and subthreshold parameters, we heavily rely
on the full $\Order(\eps^4)$ results from~\cite{Siemens:2016xxx},
where the $T$-matrix for the process $\pi N\to\pi N$ is calculated in the
small-scale expansion
\beq
  \label{eq:1}
  \eps=\left\{\frac{p}{\Lamb},\frac{\mpi}{\Lamb},\frac{\mD-\mN}{\Lamb}\right\}\qquad\text{with}\qquad \Lamb\in\{\Lambda_\chi,\mN\},
\eeq
in the HB as well as in the covariant approach. The standard on-mass-shell
renormalization scheme is employed for the leading-order LECs, where pion, nucleon, and $\Delta$ masses are denoted by
$\mpi$, $\mN$, and $\mD$, respectively, and the axial couplings
of the nucleon and nucleon--$\Delta$ transition by $\ga$ and $h_A$ (both axial couplings are renormalized at the pion vertex instead of the axial current).
After absorbing redundant contributions proportional to 
the LECs $d_{18}$ from $\Lagr_{\pi N}^{(3)}$,\footnote{In all Lagrangians, the upper index denotes the chiral order, the lower the particle content. For explicit expressions we refer to~\cite{Siemens:2016xxx}.} 
$e_{19,20,21,22,36,37,38}$ from $\Lagr_{\pi N}^{(4)}$, 
$b_{3,6}$ from $\Lagr_{\pi N\Delta}^{(2)}$, 
$c_i^\Delta$ from $\Lagr_{\pi \Delta}^{(2)}$, 
$h_i$ from $\Lagr_{\pi N\Delta}^{(3)}$,
and $k_i$ from $\Lagr_{\pi N\Delta}^{(4)}$, the $\pi N$ scattering
amplitude at $\Order(\eps^4)$ depends on the LECs 
$c_{1,2,3,4}$ from $\Lagr_{\pi N}^{(2)}$,
$d_{1+2,3,5,14-15}$ from $\Lagr_{\pi N}^{(3)}$,
$e_{14,15,16,17,18}$ from $\Lagr_{\pi N}^{(4)}$,
$h_A$ from $\Lagr_{\pi N\Delta}^{(1)}$,
$g_1$ from $\Lagr_{\pi \Delta}^{(1)}$, and
$b_{4,5}$ from $\Lagr_{\pi N\Delta}^{(2)}$. 
In the HB approach, the LECs $c_i$,
$d_i$, and $e_i$ are renormalized to
absorb UV divergent and additional decoupling-breaking pieces.
In the covariant approach, the same set of LECs is needed to cancel UV
divergences as well as decoupling- and/or power-counting-breaking pieces~\cite{Siemens:2016hdi,Siemens:2016xxx}.
In particular, both chiral amplitudes are renormalized in such a way that the
explicit difference is of higher order only, $\Order(\eps^5)$.

Employing the standard subthreshold and threshold expansion of the
$\pi N$ scattering amplitude, we calculate both sets of the respective coefficients (explicit expressions are provided as supplementary material in the form of a \textsc{Mathematica} notebook).
Furthermore, we performed a strict chiral expansion of the covariant
expressions to check that the HB expressions determined from the HB
amplitude are reproduced. In contrast to the $\Delta$-less case, where the $13$ leading
subthreshold parameters depend on $13$ $\pi\pi N N$-LECs, the
expressions in the $\Delta$-ful case depend on $4$ additional LECs from  
the $\Delta$ sector. Thus, these additional LECs cannot be extracted
by the subthreshold matching but further constraints have to be introduced. 
In particular, we assume the following conservative estimates for those particular LECs
\begin{align}
  \label{eq:2}
    h_A&=1.40\pm0.05, &b_4+b_5&=(0\pm 5)\GeV^{-1}, \notag\\
    g_1&=2.32\pm0.26, & b_4-b_5&=(0\pm 5)\GeV^{-1},
\end{align}
motivated by large-$\Nc$ considerations and, in the case of $h_A$, supplemented by phenomenology, as explained below, where the input from phenomenology allows us to reduce the 
uncertainty compared to the large-$\Nc$ prediction alone.

Given that the contributions proportional to $h_A$ already
appear at leading order, its error is most important for the final uncertainty, 
but our assignment in~\eqref{eq:2} is still reasonably conservative. 
It is consistent with the large-$\Nc$ prediction, $h_A=1.37\pm0.15$~\cite{Dashen:1993as,Dashen:1993jt},
the value extracted from the covariant $\Delta$ width at full one-loop order $h_A=1.43\pm 0.02$~\cite{Bernard:2012hb}, and
the recent extraction from $NN$ scattering by the Granada group, $h_A=1.397\pm0.009$~\cite{Perez:2014waa}, where the error refers to statistics only.
The contribution proportional to $g_1$ starts at loop level, $\Order(\eps^3)$, and its effect on the threshold and subthreshold parameters is much less relevant. 
The estimate in~\eqref{eq:2} corresponds to its large-$\Nc$ prediction, i.e.\ $g_1=9/5\,g_A$ with an $\Order(1/\Nc^2)$ error~\cite{Dashen:1993as,Dashen:1993jt}. 
The values of $h_A$ and $g_1$ are also consistent with constraint from the $\Delta$ width recently derived in~\cite{Gegelia:2016pjm}.
Finally, the LECs $b_4$ and $b_5$ only contribute at $\Order(\eps^4)$, and their impact on our results is almost negligible.
The intervals in~\eqref{eq:2} are based on a large-$\Nc$ calculation, which sets their difference and sum as  
$b_4-b_5={3 /(2\sqrt 2)}\,c_4$ and $b_4+b_5=2\sqrt 2/3\,c_{11}^\Delta$, see~\ref{sec:largeNc}. 
The value of $c_4$ in the relation for $b_4-b_5$ refers to $\Order(\eps^2)$, see Table~\ref{tab:LECs}, which corresponds to the consistent order of $c_4$ in the large-$\Nc$ relation and also avoids possible correlations with the redundant $\Delta$-LECs absorbed into the $c_i$ at higher orders, leading to an estimate of about $1\GeV^{-1}$. 
In contrast, the unknown LEC appearing in the sum, $c_{11}^\Delta$, proportional to an isotensor contribution, is fixed to zero. Choosing uncertainties generously to cover possible deviations in both cases (e.g.\ values obtained in $\pi N\to\pi\pi N$~\cite{Siemens:2014pma}), we simply vary both combinations within $\pm 5\GeV^{-1}$. 
We also checked that taking even larger intervals for these two parameters does not produce any noticeable effect in our results.  
In addition, we employ the following numerical values
for the various LECs and masses entering the leading-order
effective Lagrangian: $\mpi = 139.57\MeV$, $\Fpi =92.2\MeV$, 
$\mN= 938.27\MeV$, $\mD=1232\MeV$~\cite{Olive:2016xmw}, and $\ga=1.289$. 
The value for $\ga$ includes the Goldberger--Treiman discrepancy parameterized by $d_{18}$, 
using a $\pi N$ coupling constant $g^2/(4\pi)=13.7$~\cite{Baru:2011bw}.
We do not study the effects of the uncertainties of those quantities, which are negligible
in comparison to the other uncertainties encountered in the calculation. 

In the following, we will proceed in close analogy to~\cite{Hoferichter:2015tha}. 
The LECs $c_i$, $d_i$, and $e_i$ are matched
order-by-order to the respective subthreshold parameters, where
we employ the values determined by the RS analysis~\cite{Hoferichter:2015hva}. 
Furthermore, the full covariance matrix between the
subthreshold parameters as well as the uncertainties of the
$\Delta$-LECs in~\eqref{eq:2} are propagated by the standard
Gaussian approach into uncertainties of the extracted LECs. 
The main difference to~\cite{Hoferichter:2015tha} is the
explicit treatment of the $\Delta(1232)$ resonance in a consistent power counting up to full one-loop order.
Hence, we repeat the matching in the two HB countings already introduced in Sect.~\ref{sec:intro}, the standard one denoted by HB-$\pi N$, and the
one employed in the few-nucleon sector denoted by HB-$NN$.
In addition, we also perform the matching in a manifestly covariant
framework, both with and without explicit $\Delta$ dynamics. 
By absorbing the Goldberger--Treiman correction due to $d_{18}$, the value of $\ga$ is
slightly changed compared to~\cite{Hoferichter:2015tha}, but the difference  constitutes an $\Order(\eps^5)$ effect. 

\begin{table*}[t!]
  \centering
  \renewcommand{\arraystretch}{1.3}
  \begin{tabular}{crrrrrr}
  \toprule
  \multicolumn{1}{c}{}&\multicolumn{2}{c}{HB-$NN$} &\multicolumn{2}{c}{HB-$\pi N$}
    &\multicolumn{2}{c}{covariant}\\\midrule
    NLO&\multicolumn{1}{c}{$Q^2$}&\multicolumn{1}{c}{$\eps^2$}
    &\multicolumn{1}{c}{$Q^2$}&\multicolumn{1}{c}{$\eps^2$}
    &\multicolumn{1}{c}{$Q^2$}&\multicolumn{1}{c}{$\eps^2$}
\\\midrule
$c_1$&$-0.74(2)$&$-0.74(2)$&$-0.74(2)$&$-0.69(2)$&$-0.74(2)$&$-0.69(3)$\\ 
$c_2$&$1.81(3)$&$-0.49(17)$&$1.81(3)$&$0.81(8)$&$1.81(3)$&$0.40(10)$\\ 
$c_3$&$-3.61(5)$&$-0.65(22)$&$-3.61(5)$&$-0.44(23)$&$-3.61(5)$&$-0.49(23)$\\ 
$c_4$&$2.44(3)$&$0.96(11)$&$2.17(3)$&$0.64(11)$&$2.17(3)$&$0.64(11)$\\ \midrule
N$^2$LO&\multicolumn{1}{c}{$Q^3$}&\multicolumn{1}{c}{$\eps^3$}
    &\multicolumn{1}{c}{$Q^3$}&\multicolumn{1}{c}{$\eps^3$}
    &\multicolumn{1}{c}{$Q^3$}&\multicolumn{1}{c}{$\eps^3$}
    \\\midrule
$c_1$&$-1.08(2)$&$-1.25(3)$&$-1.08(2)$&$-1.24(3)$&$-1.00(2)$&$-1.12(3)$\\ 
$c_2$&$3.26(3)$&$1.37(16)$&$3.26(3)$&$0.79(20)$&$2.55(3)$&$1.02(12)$\\ 
$c_3$&$-5.39(5)$&$-2.41(23)$&$-5.39(5)$&$-2.49(23)$&$-4.90(5)$&$-2.27(20)$\\ 
$c_4$&$3.62(3)$&$1.66(14)$&$3.62(3)$&$1.67(14)$&$3.08(3)$&$1.21(14)$\\ 
$d_{1+2}$&$1.02(6)$&$0.11(10)$&$1.02(6)$&$-0.09(12)$&$1.78(6)$&$0.60(10)$\\ 
$d_3$&$-0.46(2)$&$-0.81(3)$&$-0.46(2)$&$-0.45(5)$&$-1.12(2)$&$-1.44(3)$\\ 
$d_5$&$0.15(5)$&$0.80(7)$&$0.15(5)$&$0.47(6)$&$-0.05(5)$&$0.28(5)$\\ 
$d_{14-15}$&$-1.85(6)$&$-1.04(12)$&$-1.85(6)$&$-0.67(14)$&$-2.27(6)$&$-0.96(12)$\\ \midrule
N$^3$LO&\multicolumn{1}{c}{$Q^4$}&\multicolumn{1}{c}{$\eps^4$}
    &\multicolumn{1}{c}{$Q^4$}&\multicolumn{1}{c}{$\eps^4$}
    &\multicolumn{1}{c}{$Q^4$}&\multicolumn{1}{c}{$\eps^4$}
   \\\midrule
$c_1$&$-1.11(3)$&$-1.11(3)$&$-1.11(3)$&$-1.11(3)$&$-1.12(3)$&$-1.10(3)$\\ 
$c_2$&$3.61(4)$&$1.52(21)$&$3.17(3)$&$1.29(18)$&$3.35(3)$&$1.20(17)$\\ 
$c_3$&$-5.60(6)$&$-1.99(30)$&$-5.67(6)$&$-2.15(29)$&$-5.70(6)$&$-2.19(28)$\\ 
$c_4$&$4.26(4)$&$1.88(19)$&$4.35(4)$&$1.94(19)$&$3.97(3)$&$1.77(17)$\\ 
$d_{1+2}$&$6.37(9)$&$1.75(42)$&$7.66(9)$&$2.95(41)$&$4.70(7)$&$1.75(22)$\\ 
$d_3$&$-9.18(9)$&$-3.61(48)$&$-10.77(10)$&$-6.02(43)$&$-5.26(5)$&$-3.24(17)$\\ 
$d_5$&$0.87(5)$&$1.52(7)$&$0.59(5)$&$1.02(6)$&$0.31(5)$&$0.65(8)$\\ 
$d_{14-15}$&$-12.56(12)$&$-4.32(79)$&$-13.44(12)$&$-5.24(76)$&$-8.84(10)$&$-3.39(53)$\\ 
$e_{14}$&$1.16(4)$&$1.67(6)$&$0.85(4)$&$1.17(6)$&$1.17(4)$&$1.31(5)$\\ 
$e_{15}$&$-2.26(6)$&$-4.91(12)$&$-0.83(6)$&$-3.38(13)$&$-2.58(7)$&$-3.07(14)$\\ 
$e_{16}$&$-0.29(3)$&$4.16(13)$&$-2.75(3)$&$2.03(24)$&$-1.77(3)$&$1.73(16)$\\ 
$e_{17}$&$-0.17(6)$&$-0.44(6)$&$0.03(6)$&$-0.37(7)$&$-0.45(6)$&$-0.51(6)$\\ 
$e_{18}$&$-3.47(5)$&$1.43(19)$&$-4.48(5)$&$0.71(23)$&$-1.68(5)$&$1.33(13)$\\ 
    \bottomrule
  \end{tabular}
  \caption{LECs extracted from fits at NLO, N$^2$LO, and N$^3$LO in
    the HB-$NN$, HB-$\pi N$, and covariant scheme with explicit $\Delta$ degrees of freedom ($\eps^n$) and in the $\Delta$-less approach ($Q^n$).
  The units of the LECs $c_i$, $d_i$, and $e_i$ are $\GeV^{-1}$, $\GeV^{-2}$, and $\GeV^{-3}$, respectively.}
\label{tab:LECs}
\end{table*}

\begin{table*}[t!]
  \centering
  \renewcommand{\arraystretch}{1.3}
  \begin{tabular}{crrrrrrr}
  \toprule
  \multicolumn{1}{c}{}&\multicolumn{2}{c}{HB-$NN$} &\multicolumn{2}{c}{HB-$\pi N$}&\multicolumn{2}{c}{covariant}&\multicolumn{1}{c}{RS}\\\midrule
  NLO&\multicolumn{1}{c}{$Q^2$}&\multicolumn{1}{c}{$\eps^2$}
    &\multicolumn{1}{c}{$Q^2$}&\multicolumn{1}{c}{$\eps^2$}
    &\multicolumn{1}{c}{$Q^2$}&\multicolumn{1}{c}{$\eps^2$}
\\\midrule
$a_{0+}^+[M_\pi^{-1}10^{-3}]$&$-14.2$&$15.5(2.6)$&$-24.0$&$-14.4(8.9)$&$-24.1$&$-7.8(6.6)$&$-0.9(1.4)$\\ 
$a_{0+}^-[M_\pi^{-1}10^{-3}]$&$79.4$&$79.4(0)$&$79.4$&$79.4(0)$&$80.1$&$81.9(1)$&$85.4(9)$\\ 
$a_{1+}^+[M_\pi^{-3}10^{-3}]$&$97.3$&$123.5(5.9)$&$103.9$&$129.2(6.2)$&$108.6$&$130.3(6.2)$&$131.2(1.7)$\\ 
$a_{1+}^-[M_\pi^{-3}10^{-3}]$&$-62.0$&$-78.6(2.1)$&$-66.5$&$-81.7(2.1)$&$-67.4$&$-83.2(2.1)$&$-80.3(1.1)$\\ 
$a_{1-}^+[M_\pi^{-3}10^{-3}]$&$-34.6$&$-48.7(3.9)$&$-47.6$&$-56.0(4.4)$&$-43.6$&$-57.3(4.5)$&$-50.9(1.9)$\\ 
$a_{1-}^-[M_\pi^{-3}10^{-3}]$&$-7.9$&$-15.0(1.9)$&$-12.5$&$-15.9(2.2)$&$-5.7$&$-14.2(2.5)$&$-9.9(1.2)$\\ 
$b_{0+}^+[M_\pi^{-3}10^{-3}]$&$-80.0$&$-50.3(2.5)$&$-70.2$&$-42.7(8.6)$&$-53.1$&$-36.3(5.5)$&$-45.0(1.0)$\\ 
$b_{0+}^-[M_\pi^{-3}10^{-3}]$&$39.7$&$39.7(0)$&$20.1$&$26.7(5)$&$11.3$&$21.7(5)$&$4.9(8)$\\ \midrule
N$^2$LO&\multicolumn{1}{c}{$Q^3$}&\multicolumn{1}{c}{$\eps^3$}
    &\multicolumn{1}{c}{$Q^3$}&\multicolumn{1}{c}{$\eps^3$}
    &\multicolumn{1}{c}{$Q^3$}&\multicolumn{1}{c}{$\eps^3$}
    \\\midrule
$a_{0+}^+[M_\pi^{-1}10^{-3}]$&$0.5$&$-12.9(6.9)$&$0.5$&$-3.5(3.6)$&$-14.8$&$1.2(5.7)$&$-0.9(1.4)$\\ 
$a_{0+}^-[M_\pi^{-1}10^{-3}]$&$92.2$&$92.7(10)$&$92.9$&$90.5(9)$&$89.9$&$81.7(1.2)$&$85.4(9)$\\ 
$a_{1+}^+[M_\pi^{-3}10^{-3}]$&$113.8$&$124.8(5.4)$&$121.7$&$125.4(5.6)$&$116.4$&$126.8(5.2)$&$131.2(1.7)$\\ 
$a_{1+}^-[M_\pi^{-3}10^{-3}]$&$-74.8$&$-77.5(2.1)$&$-75.5$&$-78.5(2.2)$&$-75.1$&$-79.5(2.1)$&$-80.3(1.1)$\\ 
$a_{1-}^+[M_\pi^{-3}10^{-3}]$&$-54.1$&$-54.4(4.1)$&$-47.0$&$-54.2(4.1)$&$-55.5$&$-54.1(3.8)$&$-50.9(1.9)$\\ 
$a_{1-}^-[M_\pi^{-3}10^{-3}]$&$-14.1$&$-13.0(2.6)$&$-2.5$&$-7.4(2.8)$&$-10.4$&$-10.0(3.0)$&$-9.9(1.2)$\\ 
$b_{0+}^+[M_\pi^{-3}10^{-3}]$&$-45.7$&$-41.2(4.5)$&$-22.1$&$-28.8(1.5)$&$-50.9$&$-29.1(2.7)$&$-45.0(1.0)$\\ 
$b_{0+}^-[M_\pi^{-3}10^{-3}]$&$35.9$&$26.4(1.0)$&$22.6$&$17.3(8)$&$21.6$&$14.3(1.5)$&$4.9(8)$\\ \midrule
N$^3$LO&\multicolumn{1}{c}{$Q^4$}&\multicolumn{1}{c}{$\eps^4$}
    &\multicolumn{1}{c}{$Q^4$}&\multicolumn{1}{c}{$\eps^4$}
    &\multicolumn{1}{c}{$Q^4$}&\multicolumn{1}{c}{$\eps^4$}
   \\\midrule
$a_{0+}^+[M_\pi^{-1}10^{-3}]$&$-1.5$&$-1.5(8.5)$&$-8.0$&$1.4(7.5)$&$-5.7$&$-0.7(6.6)$&$-0.9(1.4)$\\ 
$a_{0+}^-[M_\pi^{-1}10^{-3}]$&$68.5$&$96.3(2.0)$&$58.6$&$69.1(1.2)$&$83.8$&$83.4(1.0)$&$85.4(9)$\\ 
$a_{1+}^+[M_\pi^{-3}10^{-3}]$&$134.3$&$136.2(8.2)$&$132.1$&$135.8(7.9)$&$128.0$&$132.7(7.6)$&$131.2(1.7)$\\ 
$a_{1+}^-[M_\pi^{-3}10^{-3}]$&$-80.9$&$-80.0(3.0)$&$-90.1$&$-86.5(3.1)$&$-78.1$&$-81.1(2.1)$&$-80.3(1.1)$\\ 
$a_{1-}^+[M_\pi^{-3}10^{-3}]$&$-55.7$&$-47.2(5.0)$&$-73.7$&$-56.6(4.6)$&$-53.5$&$-51.4(4.9)$&$-50.9(1.9)$\\ 
$a_{1-}^-[M_\pi^{-3}10^{-3}]$&$-10.0$&$-6.0(2.9)$&$-23.7$&$-15.2(2.8)$&$-11.8$&$-10.3(3.9)$&$-9.9(1.2)$\\ 
$b_{0+}^+[M_\pi^{-3}10^{-3}]$&$-42.2$&$-30.8(7.9)$&$-44.5$&$-30.6(7.3)$&$-54.7$&$-33.8(6.6)$&$-45.0(1.0)$\\ 
$b_{0+}^-[M_\pi^{-3}10^{-3}]$&$-31.6$&$7.6(2.3)$&$-65.2$&$-35.0(2.3)$&$2.3$&$2.8(2.8)$&$4.9(8)$\\ \bottomrule
  \end{tabular}
  \caption{Threshold parameters predicted at next-to-leading (NLO), next-to-next-to-leading (N$^2$LO), and next-to-next-to-next-to-leading (N$^3$LO) order in
    the HB-$NN$, HB-$\pi N$, and covariant scheme with explicit $\Delta$ degrees of freedom  and in the $\Delta$-less approach, in comparison 
    to the values determined by the RS analysis.
    The orders refer to the counting in the small-scale expansion ($\eps^n$) and the ChPT expansion parameter ($Q^n$), respectively. 
    The quoted errors only cover the uncertainties propagated from the RS subthreshold parameters (and the $\Delta$ couplings where applicable), while the 
    systematic uncertainties related to the chiral expansion can be inferred by comparing the subsequent chiral orders.}
\label{tab:ThrPara}
\end{table*}

\section{Results including the $\Delta(1232)$}
\label{sec:Delta}

The extracted LECs in both HB approaches are given in Table~\ref{tab:LECs},
comparing the $\Delta$-ful and $\Delta$-less approaches. As expected~\cite{Bernard:1995dp,Bernard:1996gq,Krebs:2007rh,Fettes:2000bb,Alarcon:2012kn,Chen:2012nx,Yao:2016vbz}, 
one can observe a strong reduction of the size of the
LECs $c_i$ and $d_i$ when the $\Delta$ is considered explicitly. In
contrast, the propagated errors turn out to be somewhat larger than in the $\Delta$-less
case. Obviously this is due to the errors stemming from the additional
$\Delta$-LECs in~\eqref{eq:2}, mainly from the uncertainty in
$h_A$. In particular, the results at $\Order(\eps^3)$ look very
convincing with all LECs of natural size.
At $\Order(\eps^4)$, however, the extracted values for the LECs $d_3$ and $d_{14-15}$
still appear unnaturally large, especially in the HB-$\pi N$ counting. 
This behavior is not unexpected given that the critical combination $\ga^2(c_3-c_4)$ is still large, albeit markedly reduced,
which reflects the fact that even though in a resonance-saturation picture the $\Delta(1232)$ indeed contributes 
strongly to $c_3$ and $c_4$, additional resonances are required for a quantitative understanding~\cite{Bernard:1995dp,Bernard:1996gq}.
As for the fourth-order LECs, some $e_i$
even increase in magnitude when the $\Delta$ is included.
The correlation coefficients at each order are summarized in~\ref{sec:tables} in Table~\ref{tab:CorrHB}.

Our main results, the predictions for the eight leading threshold parameters, are collected in
Table~\ref{tab:ThrPara},
once again, comparing both HB
countings of $1/\mN$ contributions and $\Delta$-ful and $\Delta$-less approaches. 
We also show the results from the
RS analysis~\cite{Hoferichter:2015hva} as a benchmark. 
As already observed in~\cite{Hoferichter:2015tha,Hoferichter:2015hva}, the predictions
in the $\Delta$-less case do not reproduce the RS results, a deficiency that becomes most notable
in the $a_{0+}^-$ and $b_{0+}^-$ parameters, since these parameters depend most strongly 
on the critical combination of $d_i$. 
In general, the convergence is
quite poor, which is reflected by strong changes of the predicted
threshold parameters between chiral orders.
Including the $\Delta$ explicitly visibly improves this convergence pattern,
as the differences between the chiral orders are reduced and the
predictions at the highest order considered are in the same ballpark
as the RS results. 
We also show the uncertainties
propagated from the LECs taken as input in the prediction of the
threshold parameters, which prove to be of the same size as the theoretical error due to the truncation of the chiral series.
This is clearly not the case in the $\Delta$-less case, where the
statistical error is negligibly small compared to the truncation error (and therefore not displayed). 
The results are in reasonable agreement with the RS result except for $a_{0+}^-$, which is significantly 
over-predicted in the HB-$NN$ counting and strongly under-predicted in the
HB-$\pi N$ counting, and $b_{0+}^-$ in the HB-$\pi N$ case.
Finally, just considering the predictions for the  mean values, one observes
that almost all parameters deviate noticeably from the RS values, so
that agreement is only found within the relatively conservative error estimates
for the $\Delta$-LECs.

\section{Results in a covariant formulation}
\label{sec:covariant}

We start off emphasizing that based on the employed power counting, there is no \textit{a priori} argument why a manifestly covariant
scheme should give improved results compared to the HB approach.
However, since there are empirical indications that a resummation of $1/\mN$ 
corrections can lead to phenomenological improvements~\cite{Becher:2001hv,Alarcon:2012kn,Siemens:2016hdi},
we consider here the covariant analog of the HB approach discussed in the previous section.
While both HB countings of $1/\mN$ contributions yield
a consistent picture, they still display visible differences among each other and to the RS results.
A covariant approach, resumming an
infinite series of $1/\mN$ contributions, 
is not uniquely defined. In our case, we employ a covariant resummation as laid out
in~\cite{Siemens:2016xxx} and sketched in Sect.~\ref{sec:formalism}, which ensures that the differences to the HB results
start at $\Order(\eps^5)$, so that the LECs in the covariant and HB schemes can be identified with each other.
The numerical evaluation of the scalar loop functions
was done with the \textsc{LoopTools} package~\cite{Hahn:1998yk}.

The results for the extracted LECs from the matching to the
subthreshold parameters are also given in Table~\ref{tab:LECs},
comparing the case with and without explicit $\Delta$ degrees
of freedom. As can be seen, the values of the LECs in the $\Delta$-less case are
quite similar to the HB results up to order $Q^3$. At order $Q^4$, the
$c_i$ and $e_i$ are also consistent with HB, but the $d_i$ are noticeably
smaller in size. In particular the previously problematic values of 
$d_3$ and $d_{14-15}$ are reduced by roughly $50\%$ in the covariant approach.
Including the explicit $\Delta$ dynamics even further decreases those
LECs. Moreover, in the $\Delta$-ful case, the differences of the LECs between chiral orders are
reasonably small and all LECs turn out to be of natural size.
This is already a good indicator that the convergence in the covariant
case is improved compared to the HB cases. The correlation matrices
in the $\Delta$-less and $\Delta$-ful case are given in~\ref{sec:tables} in Table~\ref{tab:CorrCov}.

The predictions for the threshold
parameters in comparison to the previous HB results and the RS values are included in Table~\ref{tab:ThrPara}.
As can be seen, already the covariant $\Delta$-less results at $\Order(Q^4)$ are in
reasonable agreement with the values given by the RS
analysis. Moreover, the changes between the chiral orders are much
smaller than in both HB countings, even though the dominant error
still originates from the truncation of the chiral series. 
The convergence pattern improves further by including the $\Delta$ explicitly,
with changes between the chiral orders being small and even negligible
compared to the propagated errors stemming from the
uncertainties of the $\Delta$-LECs. It is the propagation of these uncertainties that explains the increase
of the errors at higher orders, where variations of the $h_A$ and $g_1$ central values become much more significant. 
Moreover, if we just consider the highest-order results, at $\Order(\eps^4)$, and neglect the statistical
error completely, we observe that most mean values are perfectly
consistent with the RS results, only the values of the effective
ranges $b_{0+}^\pm$ are too small in magnitude.
In conclusion, it is apparent that the results in the covariant framework present a significant
improvement over the two HB approaches.
In fact, the final uncertainties in the threshold parameters are dominated by the error estimates for $h_A$ and $g_1$. Thus, the observed agreement of the central values suggests that the errors of the $\Delta$-LECs might be overestimated, 
but a more reliable determination of those couplings is needed to draw firmer conclusions.

Beyond these empirical findings, it would be important to understand the reason for
the improvement in the covariant case. It is well-known that a covariant formulation is 
preferable in a situation where the HB expansion leads to distortions of the analytic structure,
e.g.\ in the case of anomalous thresholds~\cite{Becher:1999he,Kubis:2000zd},
but the HB formulation reproduces the analytical structure of the $\pi N$ amplitude
in the threshold and subthreshold regions correctly, where, in addition, $1/\mN$ corrections are expected to be small.
Thus, in order to further investigate the improved convergence in the
covariant approach, we have also analyzed the strict chiral expansion of the covariant expressions for the subthreshold parameters. 
We observed a very slow and oscillating convergence, 
whose origin can be traced back to the convergence pattern of nonanalytic functions in $\mpi^2$ such as
$\arctan(\mpi/\mN)$, which introduce  additional
factors of $\pi$, and especially higher-order chiral logarithms such as $\log(\mpi^2/\mN^2)$ into the chiral expansion. 
Such chiral logarithms only appear at higher chiral order, $\Order(\eps^5)$, in the covariant expressions and become
absorbed into LECs at lower chiral orders (they  would be completely absorbed into LECs in a strict HB framework).
In particular, such functions are split into an infrared singular and
regular part~\cite{Becher:1999he,Becher:2001hv} according to
  \begin{equation}
  \label{eq:3}
    \log\frac{\mpi^2}{\mN^2} 
    =\bigg[32\pi^2\bar\lambda+\log\frac{\mpi^2}{\mu^2}\bigg]
    -\bigg[32\pi^2\bar\lambda+\log\frac{\mN^2}{\mu^2}\bigg],
  \end{equation}
where the two parts can be associated with the pion and nucleon
tadpoles, respectively (with the divergent part included in $\bar\lambda$). The nucleon tadpoles are not present in the HB
amplitude at all, but already absorbed into LECs on the level of the
effective Lagrangian, whereas the pion tadpoles are absorbed by the
renormalization procedure. This implies that in the HB framework, the
LECs at higher order will receive large contributions from such chiral logarithms.
However, the scales that appear in some chiral logarithms, beyond $\Order(\eps^4)$, are in principle arbitrary, and our covariant formulation
corresponds to one admissible choice. 
Empirically, we can thus confirm that this choice allows one to remove one class of large contributions, but due 
to the lack of a power-counting argument it remains unclear
if this mechanism is universal. 
  
\section{Conclusions}
\label{sec:conclusions}

In this paper, we have studied whether threshold and subthreshold kinematics in $\pi N$ scattering
can be reconciled within ChPT by including the $\Delta(1232)$ as an explicit degree of freedom and/or using a covariant formulation.
To this end, we have performed the matching of $\pi N$ subthreshold parameters determined by Roy--Steiner equations to ChPT,
extending previous work~\cite{Hoferichter:2015tha,Hoferichter:2015hva} by including the $\Delta(1232)$ in a consistent power counting up to full one-loop order in the heavy-baryon 
as well as in a covariant framework. 
As a result, we have observed a sizable reduction of the magnitude of the extracted LECs
when the $\Delta$ is included explicitly, which, in turn, leads to an improvement of the convergence pattern in the threshold region. 
The LECs from the $\Delta$ sector, $h_A$, $g_1$, and $b_{4,5}$, have been estimated by taking into account naturalness and large-$\Nc$ constraints, and,
in the case of $h_A$, by taking into account constraints from the $\Delta$ width and $NN$ scattering. 
Based on the extracted LECs, we have calculated the eight leading threshold parameters, which, within uncertainties, become largely consistent 
with the values determined by the Roy--Steiner analysis once the $\Delta$ is included explicitly.  
Moreover, we find that the chiral convergence pattern and consistency with the threshold region improve further in
a covariant formulation. 
On a technical level, we identify terms in the covariant scheme that, once mapped onto the heavy-baryon expansion, contribute to the slow-down of the expansion,
but due to the lack of a rigorous power-counting argument it is not guaranteed that this mechanism works in general. A more profound argument
why the covariant resummation improves the phenomenological behavior would be highly desirable.

\section*{Acknowledgments}

We thank Matthias F.\ M.\ Lutz for helpful discussions, and Enrique Ruiz Arriola for useful e-mail communication. 
Financial support by
the DFG (SFB/TR 16, ``Subnuclear Structure of Matter''),  
the DOE (Grant No.\ DE-FG02-00ER41132), 
the National Science Foundation (Grant No.\ NSF PHY-1125915),
and the Swiss National Science Foundation
is gratefully acknowledged.
The work of DS was supported in part by the Ruhr University Research
School PLUS, funded by Germany's Excellence Initiative (DFG GSC 98/3).
The work of UGM was supported in part by The Chinese Academy of Sciences 
(CAS) President's International Fellowship Initiative (PIFI) grant No.\ 2015VMA076.
   
\appendix

\section{Large-$\Nc$ constraints}
\label{sec:largeNc}

In this appendix we address the derivation of the large-$\Nc$ constraints for one- and two-pion $N\Delta$ and $\Delta\Delta$ couplings that appear within our formalism.
The starting point for this analysis is the set of consistency equations derived by Dashen, Manohar, and Jenkins in~\cite{Dashen:1993as,Dashen:1993jt,Jenkins:1993af,Dashen:1993ac,Dashen:1994qi}, 
which rule the behavior of pion--baryon scattering in the large-$\Nc$ limit. 

These consistency conditions result from large-$\Nc$ QCD with nucleons interacting with a low-energy pion being an inconsistent theory: 
low-energy $\pi N$ interactions at large-$\Nc$ are dominated by the two pole-term graphs~\cite{Dashen:1993as,Dashen:1993jt}, 
leading to an overall scattering amplitude scaling with $\Nc$, which violates unitarity as well as Witten's large-$\Nc$ rules for meson--baryon scattering~\cite{Witten:1979kh}. 
This inconsistency can be cured assuming two main conditions. 
On the one hand, $\pi N$ interactions at large-$\Nc$ require the existence of an infinite tower of degenerate baryon states, 
which have to be included in the $\pi N$ scattering pole-term projection.  
On the other hand, pion--baryon axial operators (those associated with $g_A$, $h_A$, and $g_1$) commute in the large-$\Nc$ limit. 
These commutation relations already allow one to determine pion--baryon coupling relations: the Wigner--Eckart theorem expresses baryon--baryon axial matrix elements 
up to an overall unknown scale in terms of Clebsch--Gordan coefficients and reduced matrix elements, which can be computed solving the large-$\Nc$ consistency conditions~\cite{Dashen:1993as}.
In more detail, these consistency conditions imply that pion--baryon couplings can be related recursively~\cite{Jenkins:1993af}: 
the large-$\Nc$ LO contribution to $\pi N\to\pi N$ scattering occurs through a nucleon- and a $\Delta$-pole exchange, 
proportional to $g_A^2$ and $h_A^2$, respectively. Thus, the cancellation of this $\Order(\Nc)$ contribution imposes $g_A^2\sim h_A^2$, 
which fixes the $\Delta(1232)$ axial coupling up to an overall sign. Nonetheless, one also has the freedom to redefine the sign of the $\Delta$ field in $\Lagr_{\pi N\Delta}$, 
hence a positive value for $h_A$ can be picked without loss of generality. In the same way, the same cancellation in $\pi N\to \pi\Delta$ scattering requires $g_A h_A \sim h_A g_1$, 
which unambiguously relates $g_1\sim g_A$, hence fixes the sign of $g_1$ relative to $g_A$. Higher relations can be constructed proceeding similarly.

Furthermore, pion--baryon axial operators are spin-one and isospin-one tensors, and thus they also satisfy a set of commutation relations with spin and isospin generators, 
leading to an SU(4) spin-flavor contracted algebra for baryons at large-$\Nc$~\cite{Dashen:1993jt,Dashen:1994qi}. 
The irreducible representations of this contracted algebra are the solutions of the pion--baryon consistency conditions and also allow one to identify the pion coupling constant among two baryons within the same degenerate tower of $J=I=1/2,3/2,\ldots$, states in terms of an overall coupling constant~\cite{Dashen:1993jt}
\begin{align}\label{eq:piBB}
\bra{J^\prime J_3^\prime I_3^\prime} \Op^{ia} \ket{J J_3 I_3}
&=g\sqrt{{2J+1\over 2J^\prime+1}}\\\nonumber
&\times \clebsch J 1 {J^\prime} {I_3} a
{I_3^\prime} \clebsch J 1 {J^\prime} {J_3} i {J_3^\prime},
\end{align}
where $\Op^{ia}$ stands for the spatial component of the axial-current operator,\footnote{In the large-$\Nc$ limit baryon masses are $\Order(\Nc)$ whereas pion masses are $\Order(1)$. Thus, pion--baryon couplings can be studied in the rest frame of the baryon, in which the axial matrix element's time component vanishes.}  
$\bra {B^\prime}\bar\psi\gamma^i\gamma_5\tau^a\psi\ket B=\bra {B^\prime}\Op^{ia}\ket B$, and the brackets refer to the Clebsch--Gordan coefficients.
Computing these matrix elements at leading order in ChPT, \eqref{eq:piBB} allows one to identify the first relations between $\pi NN$, $\pi\Delta\Delta$, and $\pi N\Delta$ couplings
\begin{equation}\label{eq:gahAg1}
g_A=\frac{2\sqrt 2}{3}h_A=\frac{5}{9}g_1.
\end{equation}
Furthermore, the first $1/\Nc$ correction to $\Op^{ia}$ vanishes, so the relations~\eqref{eq:gahAg1} are expected to hold within a $1/\Nc^2$ uncertainty~\cite{Dashen:1993as,Dashen:1993jt,Dashen:1993ac}
\begin{equation}\label{eq:gahAg12}
h_A=\frac{3}{2\sqrt 2}g_A\left(1+\frac{ \epsilon_{h_A}}{\Nc^2}\right),\quad g_1=\frac{9}{5}g_A\left(1+\frac{\epsilon_{g_1}}{\Nc^2}\right),
\end{equation}
where $\epsilon_{h_A}$ and $\epsilon_{g_1}$ are constants of $\Order(1)$. Their values are unknown absent an explicit calculation of $1/\Nc^2$ effects,
so that for the large-$\Nc$-based error estimate for $g_1$ in~\eqref{eq:2} we put $\epsilon_{g_1}=1$.

Constraints for $\pi\pi N\Delta$ and $\pi\pi\Delta\Delta$ couplings require the analysis of two-axial-current matrix elements, 
which, in the large-$\Nc$ limit, can be written in terms of two-body spin, flavor, or axial-current operators, 
i.e.\ the generators of the SU(4) contracted algebra. Furthermore, the product of two-body operators can be expressed as a sum of a symmetric and an antisymmetric product. 
Antisymmetric combinations are directly given by the commutation relations of the SU(4) contracted algebra, 
whereas the symmetric products were worked out in~\cite{Lutz:2001yb,Lutz:2010se,Lutz:2011fe,Lutz:2014jja}.
Thus, in order to compute two-axial-current matrix elements, only one-body pion--baryon operators have to be calculated,
which can be done using a baryon state mean-field approximation~\cite{Witten:1979kh,Luty:1993fu,Lutz:2001yb,Lutz:2010se}.
This analysis was carried out in~\cite{Lutz:2001yb,Lutz:2010se}, and the outcome was matched to a three-flavor chiral-Lagrangian result, 
leading to a set of large-$\Nc$ relations of two-body counterterms. The matching of these results to our two-flavor ChPT formalism provides the relations
\begin{align}\label{eq:two-axial-current}
c_{12}^\Delta &=-c_{11}^\Delta,&b_4-b_5 &=\frac{1}{3\sqrt{2}}(9c_4-2c_4^\Delta), \notag\\
c_{13}^\Delta&=-\frac{6}{7}\left(2c_2-c_2^\Delta\right),&b_4+b_5&=\frac{2\sqrt{2}}{3}c_{11}^\Delta,
\end{align}
where, as already introduced in Sect.~\ref{sec:formalism}, the various couplings refer to the $\pi\pi NN$-, $\pi\pi\Delta\Delta$-, and $\pi\pi N\Delta$-vertex operators
according to
\beq
\label{Lagr_couplings}
\Lagr_{\pi N}^{(2)}=\sum_ic_i\Op_i^{\pi N},\quad \Lagr_{\pi \Delta}^{(2)}=\sum_ic_i^\Delta\Op_i^{\pi \Delta},\quad 
\Lagr_{\pi N \Delta}^{(2)}=\sum_ib_i\Op_i^{\pi N\Delta}.
\eeq

Nevertheless, further relations can be obtained as a direct application of the Wigner--Eckart theorem. 
Considering a large-$\Nc$ degenerate baryon spectrum of states with $I=J=1/2,3/2,\ldots$, a two-pion--baryon--baryon matrix element is given by 
\begin{align}\label{eq:twopionWE}
\bra{ J^\prime J_3^\prime I_3^\prime} \Op^{\hat J,i;\hat I,a} \ket{J J_3 I_3}
&= \Op_{\hat J,\hat I}(J,J^\prime) \sqrt{{2J+1\over 2J^\prime+1}}\\\nonumber
&\times\clebsch J {\hat I} {J^\prime} {I_3} a {I_3^\prime}
\clebsch J {\hat J} {J^\prime} {J_3} i {J_3^\prime} ,
\end{align}
where $\Op^{\hat J,i;\hat I,a}$ refers to the two-pion operator with angular momentum ${\hat J}$, isospin ${\hat I}$, and third components $i$ and $a$, respectively, 
and $\Op_{\hat J,\hat I}(J,J^\prime)$ denotes the unknown reduced matrix element. 
Thus, \eqref{eq:twopionWE} requires a spin--isospin decomposition of two-pion operators involving higher LECs.  
We perform this decomposition based on the operators in~\eqref{Lagr_couplings}. 
First, a two-pion vertex at this order can only be decomposed into partial waves with  $I,J =0,1,2$. 
Furthermore, due to Bose statistics, an isovector contribution has to be in a relative $P$-wave,
whereas isoscalar and isotensor ones can be in a relative $S$- or $D$-wave.

In the $\pi\pi NN$ sector,  $\Op_4^{\pi N}$ contributes with a vector isovector wave.  $\Op_1^{\pi N}$ and  $\Op_3^{\pi N}$ are both pure scalar isoscalar contributions, 
but $\Op_2^{\pi N}$ contributes to both isoscalar $S$- and $D$-waves. 
However, the combination that yields a pure $t$-channel $D$-wave is 
$\Op^{\pi N}_D=\Op^{\pi N}_2 - \frac{1}{6} \Op^{\pi N}_3 + \frac{1}{12} \Op^{\pi N}_1$, 
which is consistent with resonance saturation.
On the one hand, the scalar-resonance contribution to $c_{1}$ and $c_{3}$ fulfills $ c_3^S = 2 c_d/c_m c_1^S$~\cite{Bernard:1995dp}.
Furthermore, at large $\Nc$, the scalar couplings satisfy the relation $c_d=c_m$~\cite{Ledwig:2014cla}, 
so the scalar-resonance contribution to $\Op^{\pi N}_D$ vanishes, as it should. 
On the other hand, the $f_2(1270)$ resonance-exchange contribution is exactly given by the combination of operators in $\Op^{\pi N}_D$~\cite{Ecker:2007us,Dorati:2007bk}.
Thus, the scalar contribution of $\Op^{\pi N}_2$ is exactly given by the combination $\frac{1}{12}\left(2\Op^{\pi N}_3 - \Op^{\pi N}_1\right)$, 
which provides us with only two independent scalar isoscalar operators $(c_1-c_2/12)\Op^{\pi N}_1$ and  $(c_3+c_2/6)\Op^{\pi N}_3$.

Proceeding in the same way for the $\pi\pi \Delta\Delta$ sector, $c_4^\Delta$ multiplies a vector isovector operator, 
the combinations $(c_2^\Delta -2/3 c_{13}^\Delta)$ and $(c_{11}^\Delta+2/3c_{12}^\Delta)$ come together with an isoscalar $D$-wave,  $\left(c_1^\Delta-{1\over 24}c_2^\Delta-{1\over 10}c_{11}^\Delta-{1\over 15}c_{12}^\Delta+{1\over 36}c_{13}^\Delta\right)\Op^{\pi\Delta}_1$ and $\left(c_3^\Delta+{1\over 12}c_2^\Delta-{2\over 15}c_{11}^\Delta-{4\over 45}c_{12}^\Delta-{1\over 18}c_{13}^\Delta\right)\Op^{\pi\Delta}_3$ are the only two independent scalar isoscalar contributions, and $c_{12}^\Delta$ and $c_{13}^\Delta$ appear multiplying isotensor $S$- and $D$-wave terms.  

Finally, in the $\pi\pi N\Delta$ sector,  $(b_4-b_5)\Op_{4-5}^{\pi N\Delta}$ and $(b_4+b_5)\Op_{4+5}^{\pi N\Delta}$ are isovector and isotensor combinations, respectively. 

Hence, the application of \eqref{eq:twopionWE} to the scalar isoscalar operators provides the relations
\begin{align}\label{eq:I0}
c_1^\Delta-{1\over 24}c_2^\Delta-{1\over 10}c_{11}^\Delta-{1\over 15}c_{12}^\Delta+{1\over 36}c_{13}^\Delta&=c_1-{c_2\over 12}, \nonumber\\
c_3^\Delta+{1\over 12}c_2^\Delta-{2\over 15}c_{11}^\Delta-{4\over 45}c_{12}^\Delta-{1\over 18}c_{13}^\Delta&=c_3+{c_2\over6},
\end{align} 
since the normalization condition for baryon states imposes $\Op_{0,0}(J,J) =1$.

\begin{table*}[t!]
  \centering
  \renewcommand{\arraystretch}{1.3}
\begin{minipage}{0.3\linewidth}
  \begin{tabular}{crrrr}
\toprule
~~~$\eps^2$~~~&$c_1$&$c_2$&$c_3$&$c_4$\\\midrule 
$c_1$&$100$&$-17$&$25$&$-14$\\ 
$c_2$&$-2$&$100$&$-97$&$91$\\ 
$c_3$&$12$&$-99$&$100$&$-95$\\ 
$c_4$&$1$&$96$&$-95$&$100$\\ 
\bottomrule
  \end{tabular}
\end{minipage}
\begin{minipage}{0.52\linewidth}
  \begin{tabular}{crrrrrrrr}
  \toprule
$\eps^3$&$c_1$&$c_2$&$c_3$&$c_4$&$d_{1+2}$&$d_3$&$d_5$&$d_{14-15}$\\\midrule 
$c_1$&$100$&$42$&$-32$&$43$&$26$&$-34$&$12$&$-28$\\ 
$c_2$&$42$&$100$&$-99$&$97$&$91$&$-92$&$-45$&$-93$\\ 
$c_3$&$-33$&$-99$&$100$&$-95$&$-93$&$93$&$48$&$95$\\ 
$c_4$&$43$&$96$&$-95$&$100$&$86$&$-90$&$-41$&$-90$\\ 
$d_{1+2}$&$24$&$89$&$-90$&$83$&$100$&$-91$&$-62$&$-97$\\ 
$d_3$&$-21$&$-72$&$74$&$-70$&$-81$&$100$&$35$&$97$\\ 
$d_5$&$-6$&$-70$&$72$&$-66$&$-80$&$42$&$100$&$50$\\ 
$d_{14-15}$&$-24$&$-90$&$92$&$-86$&$-96$&$90$&$70$&$100$\\ 
\bottomrule
  \end{tabular}
\end{minipage}
\vskip 5pt
  \begin{tabular}{crrrrrrrrrrrrr}
  \toprule
$\eps^4$&$c_1$&$c_2$&$c_3$&$c_4$&$d_{1+2}$&$d_3$&$d_5$&$d_{14-15}$&$e_{14}$&$e_{15}$&$e_{16}$&$e_{17}$&$e_{18}$\\\midrule 
$c_1$&$100$&$-2$&$13$&$0$&$-10$&$16$&$3$&$9$&$-35$&$37$&$-14$&$-10$&$-10$\\ 
$c_2$&$-2$&$100$&$-96$&$78$&$97$&$-94$&$-41$&$-96$&$26$&$-14$&$-39$&$29$&$-54$\\ 
$c_3$&$10$&$-96$&$100$&$-84$&$-95$&$90$&$48$&$93$&$-13$&$5$&$55$&$-37$&$69$\\ 
$c_4$&$3$&$78$&$-84$&$100$&$87$&$-80$&$-45$&$-86$&$-14$&$7$&$-59$&$22$&$-80$\\ 
$d_{1+2}$&$-7$&$97$&$-96$&$88$&$100$&$-98$&$-43$&$-99$&$27$&$-22$&$-35$&$30$&$-56$\\ 
$d_3$&$11$&$-96$&$96$&$-89$&$-99$&$100$&$29$&$99$&$-41$&$30$&$20$&$-21$&$43$\\ 
$d_5$&$1$&$-48$&$59$&$-56$&$-54$&$49$&$100$&$36$&$21$&$-3$&$47$&$-38$&$48$\\ 
$d_{14-15}$&$6$&$-97$&$94$&$-87$&$-99$&$100$&$47$&$100$&$-31$&$23$&$30$&$-24$&$52$\\ 
$e_{14}$&$-34$&$11$&$8$&$-30$&$5$&$-4$&$39$&$-9$&$100$&$-86$&$74$&$-10$&$60$\\ 
$e_{15}$&$40$&$-45$&$38$&$-27$&$-50$&$46$&$18$&$48$&$-67$&$100$&$-73$&$5$&$-56$\\ 
$e_{16}$&$-28$&$-1$&$19$&$-25$&$-1$&$4$&$37$&$-2$&$93$&$-75$&$100$&$-36$&$94$\\ 
$e_{17}$&$-10$&$21$&$-28$&$11$&$23$&$-20$&$-34$&$-18$&$-8$&$-14$&$-13$&$100$&$-41$\\ 
$e_{18}$&$-14$&$-49$&$67$&$-79$&$-57$&$59$&$59$&$54$&$75$&$-24$&$76$&$-29$&$100$\\ 
\bottomrule
  \end{tabular}
  \caption{Correlation matrices at $\Order(\eps^2)$,
    $\Order(\eps^3)$, and $\Order(\eps^4)$ in the HB-$NN$ (lower triangle)
    and HB-$\pi N$ (upper triangle) counting. 
    The units of the correlation values are $10^{-2}$.}
\label{tab:CorrHB}
\end{table*}

In the same way, for the vector isovector operators one finds the relations
\begin{align}\label{eq:I1}
c_4^{\Delta}&=\frac{9}{5}\frac{\Op_{1,1}(3/2,3/2)}{\Op_{1,1}(1/2,1/2)} c_4, \nonumber\\
b_4-b_5&={3 \sqrt2\over2}\frac{\Op_{1,1}(1/2,3/2)}{\Op_{1,1}(1/2,1/2)} c_4,
\end{align}
and for the isotensor 
\begin{align}\label{eq:I2}
b_4+b_5={1\over 6\sqrt{2}}\frac{\Op_{2,2}(1/2,3/2)}{\Op_{2,2}(3/2,3/2)}c_{12}^\Delta. 
\end{align}
In the diagonal case $J=J'$ (and within the same $I=J$ spectrum) the reduced matrix element factorizes 
into standard angular-momentum reduced matrix elements $X_{\hat J}$~\cite{Racah:1942gsc,Biedenharn:1981er}
\beq
\Op_{\hat J,\hat I}(J,J)=\frac{X_{\hat J}(J)}{(2J+1)^{3/2}} \frac{X_{\hat I}(I)}{(2I+1)^{3/2}},
\eeq
which due to $X_1(J)=\sqrt{J(J+1)(2J+1)}$ produces
\beq
\Op_{1,1}(J,J)=\frac{J(J+1)}{(2J+1)^2}.
\eeq
The combination of the two-axial-current matrix element constraints in \eqref{eq:two-axial-current} with those in \eqref{eq:I0}, \eqref{eq:I1}, and \eqref{eq:I2} then provides the relations
\begin{align}\label{prefinal}
b_4-b_5&={3\over 2\sqrt 2} c_4, &
b_4+b_5&={2\sqrt 2\over 3}c_{11}^\Delta, \nonumber\\
{c_2^\Delta\over 2}&=c_2-28(c_1-c_1^\Delta)-{14\over 15}c_{11}^\Delta, &
c_4^\Delta&={9\over 4} c_4, \nonumber\\
c_3^\Delta&=c_3+2(c_1-c_1^\Delta)+{1\over 9}c_{11}^\Delta,&
c_{12}^{\Delta}&=-c_{11}^\Delta,\nonumber\\
{c_{13}^\Delta\over 8}&=-6(c_1-c_1^\Delta)-{1\over 5}c_{11}^\Delta.
\end{align}

The constraints derived from the two-axial-vector currents involve all SU(2) LECs but $c_1$ and $c_1^\Delta$, which appear multiplying pure explicit-symmetry-breaking terms. 
Since the nucleon and $\Delta(1232)$ become degenerate at large $\Nc$, it is natural to assume that these explicit-symmetry-breaking LECs must be equal in this limit, $c_1^\Delta=c_1$. 
With this last assumption, \eqref{prefinal} simplifies accordingly, dropping all the contributions $\propto(c_1-c_1^\Delta)$.

\begin{table*}[t!]
  \centering
  \renewcommand{\arraystretch}{1.3}
\begin{minipage}{0.3\linewidth}
  \begin{tabular}{crrrr}\toprule
NLO&$c_1$&$c_2$&$c_3$&$c_4$\\\midrule 
$c_1$&$100$&$-17$&$26$&$-14$\\ 
$c_2$&$-9$&$100$&$-99$&$94$\\ 
$c_3$&$50$&$-85$&$100$&$-95$\\ 
$c_4$&$4$&$6$&$-4$&$100$\\ 
\bottomrule
  \end{tabular}
\end{minipage}
\begin{minipage}{0.52\linewidth}
  \begin{tabular}{crrrrrrrr}\toprule
N$^2$LO&$c_1$&$c_2$&$c_3$&$c_4$&$d_{1+2}$&$d_3$&$d_5$&$d_{14-15}$\\\midrule 
$c_1$&$100$&$24$&$-12$&$28$&$19$&$23$&$10$&$-17$\\ 
$c_2$&$-9$&$100$&$-99$&$96$&$88$&$46$&$-44$&$-89$\\ 
$c_3$&$50$&$-85$&$100$&$-95$&$-88$&$-43$&$46$&$91$\\ 
$c_4$&$4$&$6$&$-4$&$100$&$82$&$52$&$-40$&$-86$\\ 
$d_{1+2}$&$-27$&$64$&$-71$&$9$&$100$&$18$&$-63$&$-96$\\ 
$d_3$&$13$&$-40$&$55$&$-15$&$-61$&$100$&$-27$&$-13$\\ 
$d_5$&$37$&$-24$&$35$&$-2$&$-58$&$-6$&$100$&$50$\\ 
$d_{14-15}$&$31$&$-48$&$67$&$-10$&$-86$&$86$&$31$&$100$\\ 
\bottomrule
  \end{tabular}
\end{minipage}
\vskip 5pt
  \begin{tabular}{crrrrrrrrrrrrr}\toprule
N$^3$LO&$c_1$&$c_2$&$c_3$&$c_4$&$d_{1+2}$&$d_3$&$d_5$&$d_{14-15}$&$e_{14}$&$e_{15}$&$e_{16}$&$e_{17}$&$e_{18}$\\\midrule
$c_1$&$100$&$-3$&$16$&$-3$&$-16$&$22$&$14$&$11$&$-41$&$29$&$-20$&$-13$&$-11$\\ 
$c_2$&$15$&$100$&$-99$&$85$&$75$&$-95$&$2$&$-96$&$-5$&$-75$&$-36$&$19$&$-51$\\ 
$c_3$&$58$&$-67$&$100$&$-85$&$-80$&$98$&$3$&$95$&$2$&$76$&$38$&$-22$&$53$\\ 
$c_4$&$6$&$1$&$4$&$100$&$69$&$-89$&$1$&$-81$&$-22$&$-62$&$-47$&$-2$&$-59$\\ 
$d_{1+2}$&$-36$&$70$&$-82$&$9$&$100$&$-75$&$-54$&$-60$&$-30$&$-49$&$-67$&$34$&$-74$\\ 
$d_3$&$59$&$-44$&$88$&$-24$&$-79$&$100$&$-5$&$95$&$-8$&$80$&$29$&$-16$&$47$\\ 
$d_5$&$21$&$-25$&$27$&$4$&$-51$&$12$&$100$&$-20$&$40$&$-12$&$52$&$-28$&$41$\\ 
$d_{14-15}$&$45$&$-58$&$85$&$-23$&$-92$&$94$&$28$&$100$&$-15$&$81$&$15$&$-12$&$34$\\ 
$e_{14}$&$-59$&$59$&$-97$&$-3$&$85$&$-93$&$-25$&$-91$&$100$&$-51$&$88$&$3$&$77$\\ 
$e_{15}$&$36$&$-62$&$72$&$5$&$-82$&$55$&$45$&$69$&$-73$&$100$&$-22$&$-17$&$-7$\\ 
$e_{16}$&$-11$&$60$&$-51$&$-20$&$71$&$-34$&$-33$&$-53$&$55$&$-92$&$100$&$-12$&$95$\\ 
$e_{17}$&$-12$&$20$&$-24$&$-64$&$29$&$-7$&$-20$&$-7$&$23$&$-26$&$35$&$100$&$-27$\\ 
$e_{18}$&$-23$&$57$&$-56$&$-39$&$52$&$-24$&$-37$&$-40$&$52$&$-74$&$72$&$5$&$100$\\ 
\bottomrule
  \end{tabular}
  \caption{Correlation matrices at order NLO, N$^2$LO, and N$^3$LO in the
    covariant approach with (upper triangle) and without (lower triangle)
    explicit $\Delta$ degrees of freedom. 
    The units of the correlation values are $10^{-2}$.}
\label{tab:CorrCov}
\end{table*}

\section{Correlation coefficients}
\label{sec:tables}

In this appendix we collect the correlation coefficients of the LECs for the different schemes and chiral orders, see Tables~\ref{tab:CorrHB} (HB) and~\ref{tab:CorrCov} (covariant).

\end{document}